\documentclass[12pt]{article}
\usepackage{amsmath}
\usepackage{amssymb}
\usepackage{epsfig}
\usepackage{yf}

\newcommand{\comment}[1]{}
\newcommand{\Journal}[5]{#1~#2~\textbf{#3}, #5 (#4)}

\begin{document}
\title{$\sigma$-meson in hot and dense matter}
\author{M. K. Volkov, A.E. Radzhabov, N. L. Russakovich\\[5mm]
\itshape Joint Institute for Nuclear Research, Dubna, Russia \\
}
\date{}
\maketitle \Abstract{\large An important role of the scalar-isoscalar $\sigma$-meson in the
low-energy physics is discussed. The behavior of the $\sigma$-meson in the hot and dense
medium is studied. It is shown that in the vicinity of the critical values of the
temperature($T$) and the chemical potential($\mu$) the $\sigma$-meson can become a sharp
resonance. This effect can lead to a strong enhancement of the processes
$\pi\pi\rightarrow\gamma\gamma$ and $\pi\pi\rightarrow\pi\pi$ near the two-pion threshold.
Experimental observation of this phenomenon can be interpreted as a signal of approaching the
domain where the chiral symmetry restoration and the phase transition of hadron matter into
quark-gluon plasma take place.}

\clearpage

\section{Introduction}
In the last years, the problem of studying the scalar-isoscalar $\sigma$-meson properties has
attracted great attention of many authors \cite{jws,fws}. The subjects of investigations are
internal properties of the $\sigma$-meson and its role as an intermediate particle in various
processes both in vacuum and in hot and dense matter. The latter problem is especially
important, because many experiments on heavy ion collision are performed and planned (CERN,
Brookhaven, DESY, Darmstadt). One of the aims of those experiments is to study the problem of
phase transition of hadron matter into quark-gluon plasma. Special workshops were dedicated to
the study of the $\sigma$-meson properties, in Japan, June 2000 ~\cite{jws} and in France,
September 2001 \cite{fws}. Very interesting talks on this topic were given by T. Kunihiro
\cite{kunih1,kunih2,kunihpp}. In this article, we will follow these reports.However, here we
will use the results obtained only in our previous works.

Let us start with the experimental status of the $\sigma$-meson. The experimental value of the
$\sigma$-meson mass is not accurately determined and lies in a wide interval\cite{pdg}:
\begin{equation}
M_{\sigma}= 400 - 1200MeV
\end{equation}
It is explained by large values of the decay width of this meson into two pions\footnote{Note
that for a long time the information about the $\sigma$- meson was absent in PDG and only in
1998 it appeared}\cite{pdg}:
\begin{equation}
\Gamma_{\sigma}=600 - 1000MeV    \label{expgsig}
\end{equation}

However, at high a temperature and density $M_\sigma$ can become lower than the mass of two
pions. Therefore, the decay $\sigma\rightarrow\pi\pi$ is closed, and the $\sigma$-meson
becomes a stable particle. As a result, the $\sigma$-meson can give sharp resonance when it
participates in processes as an intermediate state.

Now let us describe the theoretical status of the $\sigma$-meson. The $\sigma$-meson is a
chiral partner of the $\pi$-meson in different linear $SU(2) \times SU(2)$ $\sigma$-models
\cite{Alfaro}. On the other hand, it is a scalar-isoscalar singlet in a $U(3) \times U(3)$
symmetric quark model of the Nambu-Jona-Lasinio(NJL) type \cite{echaya86}.

In the NJL model, the $\sigma$-meson mass can be expressed via the pion mass $M_{\pi}$ and the
constituent quark mass $m$ \footnote{We would like to notice that for a more accurate
description of the mass spectra of scalar mesons it is necessary to take into account the
singlet-octet mixing of scalar-isoscalar mesons with each other and with the scalar glueball
\cite{ejpa,nuovo}.}
\begin{eqnarray}
M_{\sigma}^2=M_{\pi}^2+4m^2  \label{msig}
\end{eqnarray}
Here we assume that the constituent masses of up and down quarks are equal to each other $m_u
\approx m_d = m$.

Formula (\ref{msig}) plays a very important role for the description of different processes in
hot and dense matter where the $\sigma$-meson participates as an intermediate particle.
Indeed, from (\ref{msig}) it follows that in vacuum $M_\sigma$ is larger than $M_\pi$ because
$m \approx 280 MeV, M_\pi \approx 140 MeV, M_\sigma \approx 580 MeV$\cite{echaya86}. A
different situation occurs in hot and dense matter in the vicinity of the critical values of
the temperature and chemical potential, where $m \rightarrow m_0 \approx 0$ and $M_\sigma
\rightarrow M_\pi$. This corresponds to the restoration of the chiral symmetry. As will be
shown below, this behavior of $M_\sigma$ can lead to the resonant enhancement of some
processes where the $\sigma$-meson participates as an intermediate particle (for example $\pi
\pi \rightarrow \gamma \gamma$, $\pi \pi \rightarrow \pi \pi$). Observation of such effects,
for instance in heavy ion collisions, could indicate approaching the domain of $T$, $\mu$
values where the phase transition from hadron matter into quark-gluon plasma appears. The
possibility of such a phase transition and the chiral symmetry restoration is a subject of the
intensive investigation at present.

The paper is organized is follows. In the next section we demonstrate the important role of
the $\sigma$-meson for the correct description of different processes in vacuum. In sec.3 we
compare the behavior of the $\sigma$-meson propagator in vacuum and hot dense matter. We show
that in hot and dense matter the $\sigma$-meson propagator can be a sharp resonance. Processes
$\pi^+\pi^-\rightarrow \gamma\gamma$, $\pi^0\pi^0\rightarrow \gamma\gamma$ in vacuum and hot
dense matter are investigated in sec.4. Noticeable enhancement of these cross-sections near
the two-pion threshold in the vicinity of the critical $T$, $\mu$ values  is found. In the
last section a short discussion of the results is given. Theoretical and experimental results
concerning $\pi\pi$-scattering process in hot and dense matter are discussed.

\section{$\sigma$-meson in vacuum}

Before describing the $\sigma$-meson properties in hot and dense matter, we show a very
important role of the $\sigma$-meson in a set of processes going in vacuum. Let us consider
here some of them: $\pi \pi$-scattering, $\pi \pi \rightarrow \gamma \gamma$, the rule $\Delta
I=1/2$(where $I$ is the isospin of the meson system) in kaon decays and the calculation of the
pion-nucleon $\Sigma$-term in $\pi$-nucleon scattering. We give only a qualitative picture of
these processes. Details can be found in the original works
\cite{pppp1,VKur,VKur2,kspp1,pns1,actaphys}.

Let us start with $\pi \pi$-scattering. In the $SU(2)\times SU(2)$ chiral NJL model, this
process can be described by the lagrangian\footnote{For simplicity, we do not take into
account $\pi-a_1$ transitions($a_1$ is the axial-vector meson) }:
\begin{eqnarray}
L(q,\bar{q},\sigma,\pi)=\bar{q}(x)\left(i\hat{\partial}-M+g\left(\sigma(x)+i\gamma^5
\pmb{\tau}\pmb{\pi}(x) \right)\right) q(x) \label{qlag}
\end{eqnarray}
where $\bar{q}(x)=(\bar{u}(x),\bar{d}(x))$ is the quark field, $\sigma(x)$,$\pi(x)$ are the
$\sigma$,$\pi$ meson fields, $M$ is the diagonal mass matrix of the constituent quarks,
$\pmb{\tau}$ are the Pauli matrices, $g$ is the quark-meson strong coupling constant
\begin{eqnarray}
g=(4I_2)^{-\frac{1}{2}}
\end{eqnarray}
where $I_2$ is a logarithmical divergent integral that appears in the quark loop, $I_n$(n=1,2)
in the Euclidean metric is equal to
\begin{eqnarray}
I_n= \frac{N_c}{(2\pi)^4}\int d^4_E k\frac{\Theta(\Lambda^2-k^2)}{(m^2+k^2)^n} \label{In}
\end{eqnarray}
where $\Lambda$ is a cut-off parameter($\Lambda=1.25 GeV$)\cite{echaya86}.

\begin{figure}[h]
\begin{center}
\epsfig{figure=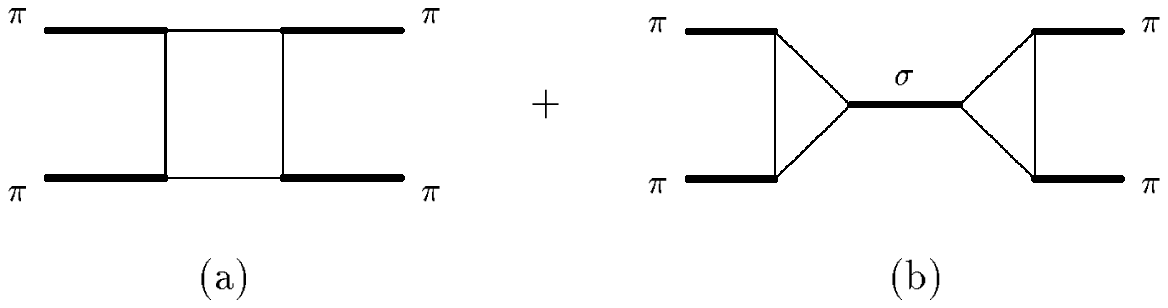,scale=1.2} \caption{} \label{pipi}
\end{center}
\end{figure}

Diagrams describing $\pi \pi$-scattering are given in Fig.\ref{pipi}. Then using lagrangian
(\ref{qlag}) we obtain the following expression for the amplitude
$A_{\pi\pi}$\cite{echaya86,pppp1}
\begin{eqnarray}
A_{\pi\pi}=-4{g^2} + \frac{(4mg)^2}{M^2_{\sigma}-s} = \frac{s-M^2_\pi}{F^2_\pi},
\label{Api}
\end{eqnarray}
where $F_{\pi}=93MeV$ is the pion weak decay constant and $s=(p_1+p_2)^2$, $p_1$ and $p_2$ are
the momenta of the incoming pions.

The final expression for $A_{\pi\pi}$ is the famous formula describing the $\pi\pi$-scattering
amplitude at low energies. This formula was first obtained by Weinberg in the sixties years of
the last century. It was one of the basic formulas demonstrating the chiral symmetry of
strong interaction\cite{Alfaro}. The following relations were used to derive this formula:\\
1. The  Goldberger- Treiman identity $g=m/F_\pi$ \\
2. Formula for the $\sigma$-meson mass (\ref{msig}). \\
3. The amplitude for the decay $\sigma \rightarrow \pi^+ \pi^-$ $A_{\sigma \rightarrow \pi^+
\pi^-}=4mg.$ \\
We can see from eq.(\ref{Api}) that the constant part of the $\sigma$ pole diagram cancels the
contribution of the box diagram. The remaining part of the $\sigma$-pole diagram determines
the $s$-dependence of the $\pi \pi$-scattering amplitude in agreement with the chiral symmetry
requirements \cite{Alfaro}.

The $\sigma$-pole diagram plays an important role for describing the polarizability of the
pion which is a significant characteristic of its electromagnetic
structure\cite{echaya86,pionpolar,pionpola}. The $\sigma$-meson is also necessary for
description of the processes $\pi \pi \rightarrow \gamma \gamma$ and $\gamma\pi \rightarrow
\gamma\pi$ in hot and dense matter \cite{VKur,VKur2}. Below we will discuss the process $\pi
\pi \rightarrow \gamma \gamma$ in detail.

The famous rule $\Delta I=1/2$ is connected with an experimentally observable enhancement of
the decay $K_S \rightarrow \pi \pi$ as compared with kaon decays with $\Delta I=3/2$. This
effect can be explained by the presence of the channel with the intermediate $\sigma$-meson
(see Fig.\ref{ksdiag}) in this process. Indeed, the $\sigma$-pole diagram in the process $K_S
\rightarrow \pi \pi$ leads to the appearance of a resonance factor in this amplitude. This
factor takes the form
\begin{equation}
\frac{1}{M^2_{\sigma}-M^2_{K_S}-iM_{\sigma}\Gamma_{\sigma}}  \label{resf}
\end{equation}
where $\Gamma_\sigma$ is the decay width of the $\sigma$ meson. The kaon mass $M_{K_S}$ is
close to the $\sigma$-meson mass. That gives a noticeable enhancement of this channel as
compared with the channels with $\Delta I=3/2$ where the $\sigma$-pole diagram cannot exist.
Then pions are emitted directly from quark loops containing a weak vertex \cite{kspp1}.

\begin{figure}[h]
\begin{center}
 \epsfig{figure=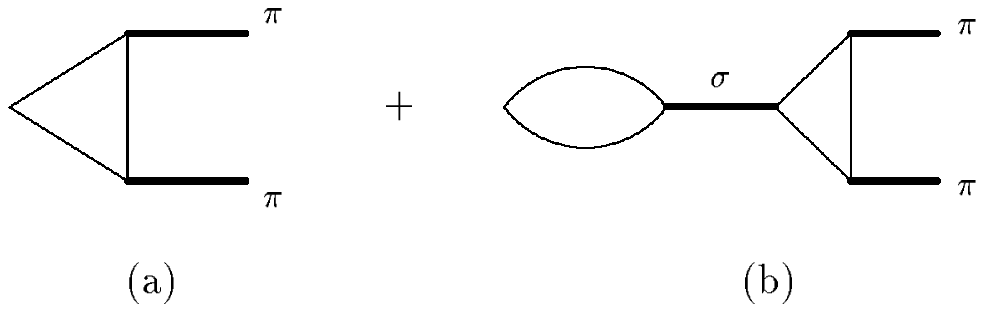,scale=1.2} \caption{} \label{spp}
\end{center}
\end{figure}

We would also like to emphasize an important role of the $\sigma$-pole diagram in the
calculation of the pion-nucleon $\Sigma$-term \cite{pns1,actaphys}. The value of the
$\Sigma$-term is determined by diagrams in the Fig.\ref{spp}. These diagrams lead to the
following expression(see \cite{actaphys})
\begin{eqnarray}
\langle \pi^+(0)\vert {\bar u}u + {\bar d}d\vert \pi^+(0)\rangle = 4 m [ 1 + \left(
I_1-2m^2I_2 \right) \frac{8 g^2}{M_\sigma^2}] \label{sigm}
\end{eqnarray}
In this formula the first term corresponds to the triangle quark diagram (Fig.\ref{spp}a), and
the second term corresponds to the $\sigma$-pole diagram (Fig.\ref{spp}b).

Taking into account the relations $g=(4I_2)^{-\frac{1}{2}}=m/F_\pi$ and $M_\sigma \approx 2m$,
we see that the first term is cancelled by the part of the second term containing $I_2$. The
remaining part of eq.(\ref{sigm}) takes the form \footnote{Here we neglect the momentum and
mass of the pion and use the formula
\begin{eqnarray}
I_1=\frac{3}{(4\pi)^2}\left(\Lambda^2 -m^2 \ln ( \frac{\Lambda^2}{m^2}+1)\right) = 0.025 GeV^2
\nonumber
\end{eqnarray} }
\begin{eqnarray}
\langle \pi^+(0)\vert {\bar u}u + {\bar d}d\vert \pi^+(0)\rangle \approx 4 m \cdot \left[ 2I_1
\frac{g^2}{m^2} = 2 \frac{I_1}{F_\pi^2}\approx 5.8\right]
\end{eqnarray}
It is easy to see that the contribution from the sum of diagrams \ref{spp}a and \ref{spp}b is
$5.8$ times as large as the contribution from diagram \ref{spp}a. As a result, in our work
\cite{pns1} for $\pi-N$ $\Sigma$ term was obtained $\Sigma_{\pi N}=50 \pm 10 MeV$. Notice that
in \cite{actaref1,actaref2,actaref3}, where the $\pi$-N $\Sigma$-term was calculated in the
framework of nonlinear chiral models, the authors obtained a small value for $\Sigma$-term
because they did not take into account the contribution from the scalar $\sigma$-meson.

\section{$\sigma$-meson propagator in vacuum and hot dense matter}

Up to now we have considered the above-mentioned processes only in vacuum. Now let us study
the $\sigma$-meson properties in hot and dense matter. It is especially interesting to
investigate the behavior of the $\sigma$-meson propagator for the process
$\pi\pi\rightarrow\pi\pi$, $\pi\pi\rightarrow\gamma\gamma$
\begin{equation}
\Delta_{\sigma}(s)=\frac{1}{M^2_{\sigma}-s-iM_{\sigma}\Gamma_{\sigma}(s)} \label{prop}
\end{equation}
where $s={(p_{\pi_1}+p_{\pi_2})}^2$, $p_1$ and $p_2$ are the momenta of the incoming pions.The
decay width $\Gamma_\sigma$ in the NJL model has the form
\begin{equation}
\Gamma_{\sigma}(s)=\frac{3m^4}{2\pi M_\sigma F^2_{\pi}}\sqrt{1-\frac{4M^2_{\pi}} {s}}
\label{gsig}
\end{equation}

In vacuum, when $T=0$ and $\mu=0$ and the constituent quark mass is $m=280MeV$, we can
consider two extreme cases:
\begin{enumerate}
\item
{$s \approx M^2_{\sigma}$. In this case the real part of the denominator of
$\Delta_{\sigma}(s)$ (\ref{prop}) is equal to zero but the imaginary part is large:
$M_{\sigma}\Gamma_{\sigma} \approx 0.3 GeV^2 $}
\item
{$s \approx 4M^2_{\pi}$. In this case the imaginary part of the denominator is close to zero,
however its real part is large $ M^2_{\sigma}-4M^2_\pi \approx 0.25 GeV^2$ }
\end{enumerate}

Therefore, in vacuum the real and imaginary parts of the denominator cannot be close to zero
simultaneously, and the $\sigma$-pole diagram cannot give a sharp resonance in the whole
domain of energy.

A more interesting situation can arise in hot and dense medium. The constituent quark mass
decreases and the pion mass slightly increases with increasing $T$ and $\mu$. Therefore the
case that $4 m^2 \approx 3M_{\pi}^2$ is possible . Then, if we consider the above-mentioned
processes near the two-pion threshold $s=4 M^2_{\pi}(1+\epsilon)$ ($\epsilon \ll 1$), we can
see that the real and imaginary parts of the $\Delta_{\sigma}(s)$ denominator become very
small simultaneously
\begin{eqnarray}
Re\left(\Delta_{\sigma}(s)^{-1}\right)&=& M_\sigma^2-s=4 m^2-3M_{\pi}^2-4M_{\pi}^2\epsilon
\approx - 4M_{\pi}^2\epsilon \\
Im\left(\Delta_{\sigma}(s)^{-1}\right)&=&-M_\sigma \Gamma_\sigma=-\frac{3m^4}{2\pi
F_{\pi}^2}\sqrt{\frac{\epsilon}{1+\epsilon}} \approx -\frac{3m^4}{2\pi F_{\pi}^2}
\sqrt{\epsilon}
\end{eqnarray}
As a result, the propagator takes the form
\begin{eqnarray}
\Delta_{\sigma} \approx \frac{1}{- 4M_{\pi}^2\epsilon-i \frac{3m^4}{2\pi
F_{\pi}^2}\sqrt{\epsilon}}
\end{eqnarray}
The formula shows that in hot and dense matter in the vicinity of the critical $T$ and $\mu$
values, the $\sigma$-meson propagator can become a sharp resonance. This leads to a noticeable
enhancement of processes where the $\sigma$-meson participates as an intermediate particle.

In the next section we demonstrate this effect on the basis of the process $\pi \pi
\rightarrow \gamma \gamma$ following the papers \cite{VKur,VKur2}.

\section{Process $\pi \pi \rightarrow \gamma \gamma$ in hot dense matter}

\begin{figure}[h]
\begin{center}
\epsfig{figure=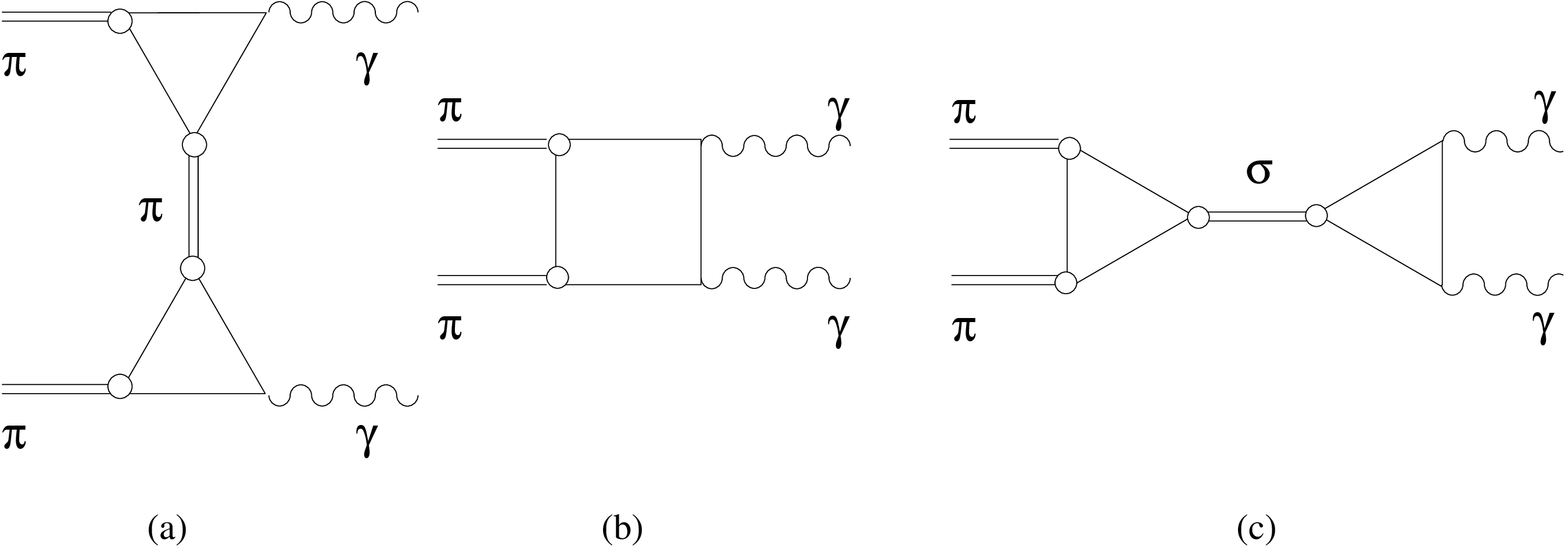,height=13cm,angle=-90} \caption{} \label{2p2g} 
\end{center}
\end{figure}

To describe the processes $\pi^+ \pi^- \rightarrow \gamma \gamma$ and $\pi^0 \pi^0 \rightarrow
\gamma \gamma$ it is necessary to consider quark loop diagrams of three types (see
Fig.\ref{2p2g}).

\begin{figure}[h]
\begin{center}
\epsfig{figure=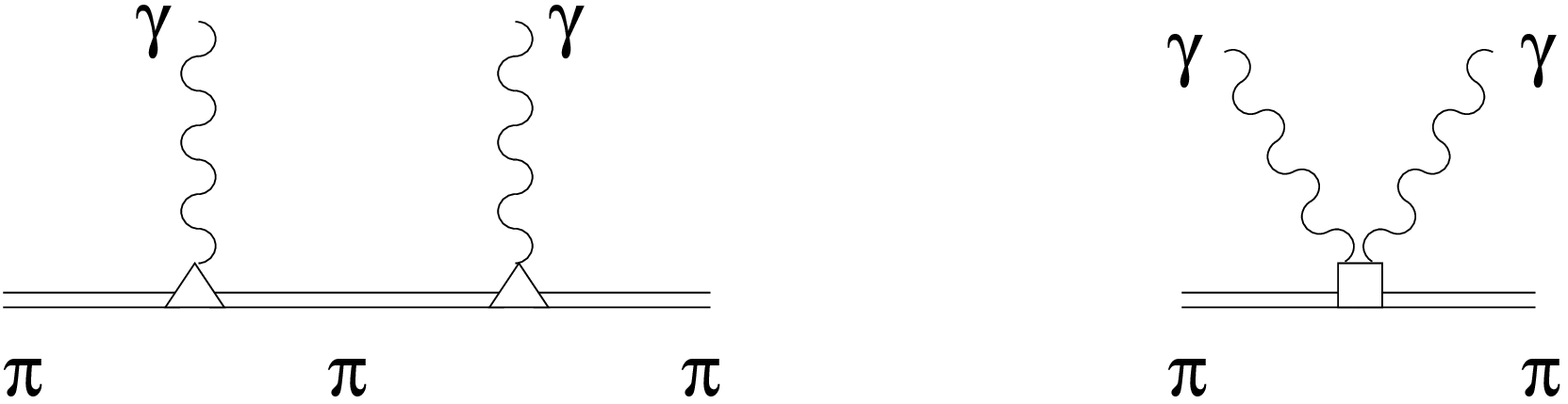,height=12cm,angle=-90} \caption{} \label{2p2gborn}
\end{center}
\end{figure}

Diagram \ref{2p2g}a exists only for charged pions. Diagrams \ref{2p2g}a and \ref{2p2g}b define
the Born terms in a local approximation(see Fig.\ref{2p2gborn}). In this approximation, only
divergent parts of quark diagrams are considered. As a result, the lagrangian for the
photon-meson vertices
\begin{equation}
L^{Born}= ieA_\mu \left[ \pi^- \partial_\mu \pi^+ - \pi^+ \partial_\mu \pi^- \right] + A_\mu^2
\pi^+ \pi^-
\end{equation}
is obtained in this approximation. In the next $k^2$-approximation only diagrams \ref{2p2g}b
and \ref{2p2g}c give nontrivial contributions. The lagrangians corresponding to the vertices
$\pi\pi\rightarrow\gamma\gamma$ and $\sigma\rightarrow \gamma\gamma$ take the form
\begin{eqnarray}
L^{box} &=& \frac{\alpha}{18 \pi F_\pi^2} \left[ \pi^+ \pi^- + 5 \pi^0 \pi^0 \right] F_{\mu \nu}^2\\
L^{\sigma \rightarrow \gamma \gamma}&=& \frac{5 \alpha}{9 \pi F_\pi}\sigma F_{\mu \nu}^2 ,
\end{eqnarray}
here we use the notation $\alpha = \frac{e^2}{4 \pi} \approx \frac{1}{137}$($e$ is the e.m.
charge), $F_{\mu \nu} = \partial_\mu A_\nu - \partial_\nu A_\mu$. The lagrangian describing
the vertex $\sigma \rightarrow \pi\pi$ has the form
\begin{eqnarray}
L^{\sigma \pi \pi} &=& 2m g \sigma \pmb{\pi}^2
\end{eqnarray}

These lagrangians allow us to define the total amplitude describing processes $\pi \pi
\rightarrow \gamma \gamma$\footnote{Let us note that in the Born approximation only diagrams
\ref{2p2g}a and \ref{2p2g}b together give a gauge-invariant expression for amplitudes. In the
$k^2$-approximation, diagrams \ref{2p2g}b and \ref{2p2g}c give a gauge-invariant expression
separately}
\begin{eqnarray}
T^{\mu \nu}(s)&=& T^{\mu \nu}_{Born}(s)+T^{\mu \nu}_{k^2}(s) \\
T^{\mu \nu}_{Born}(s)&=&2 e^2 [g^{\mu\nu}-\frac{p_1^\mu p_2^\nu}{p_1k_1}-\frac{p_2^\mu p_1^\nu}{p_2k_1}] \\
T^{\mu \nu}_{k^2}(s)&=&e^2 A(s)[g^{\mu \nu} k_1 k_2 - k_1^\mu k_2^\nu] \\
A(s)&=&\frac{1}{(6 \pi F_\pi)^2} \left[ \frac{40 m^2}{M_\sigma^2-s-iM_\sigma \Gamma_\sigma} -
1 \right]
\end{eqnarray}
where $s=(p_1+p_2)^2$, $p_i$ and $k_i$ are the momenta of the pions and photons, respectively,
$\Gamma_{\sigma}$ is the decay width of the $\sigma$-meson (see (\ref{gsig})).

For the process $\pi^0\pi^0\rightarrow\gamma\gamma$ the Born term is absent. The contribution
of the box diagram increases 10 times\cite{echaya86}.

The cross-section of the process $\pi^+ \pi^- \rightarrow \gamma \gamma$ consists of three
parts
\begin{eqnarray}
\sigma_{\pi^+ \pi^- \rightarrow \gamma \gamma} &=& \frac{\pi \alpha^2}{4s \kappa }\left[
\tilde{\sigma}_1 + \tilde{\sigma}_2 + \tilde{\sigma}_3\right]
\end{eqnarray}
where
\begin{eqnarray}
\kappa=\sqrt{1-\frac{4M_\pi^2}{s}}
\end{eqnarray}

Here $\tilde{\sigma}_1$ corresponds to the Born term, $\tilde{\sigma}_3$ corresponds to
contributions from the $\sigma$-pole and box diagrams, and $\tilde{\sigma}_2$ is the
interference term of the Born and $k^2$-approximation contributions. For the neutral pion we
have only the $\tilde{\sigma}_3$ term, where box the contribution is ten times larger than in
the case with charge pions. Further we will consider these processes near the two-pion
threshold. Variables $s$ and $\kappa$ in this domain take the form
\begin{eqnarray}
s=4M_\pi^2(1+\epsilon),\kappa=\sqrt{\frac{\epsilon}{1+\epsilon}}
\end{eqnarray}
where $\epsilon \ll 1$. Then for $\tilde{\sigma}_1$,$\tilde{\sigma}_2$,$\tilde{\sigma}_3$ we
have
\begin{eqnarray}
\tilde{\sigma}_1&=& 16\left[ 2 - \kappa^2 -\frac{1-\kappa^4}{2 \kappa}\ln\frac{1+\kappa}{1-\kappa}\right] \nonumber \\
\tilde{\sigma}_2&=& 4sReA(s)\frac{1-\kappa^2}{\kappa}\ln\frac{1+\kappa}{1-\kappa} \label{partsigma}\\
\tilde{\sigma}_3&=& s^2 |A(s)|^2 \nonumber
\end{eqnarray}
The $\tilde{\sigma}_3$ term has the form
\begin{eqnarray}
 \tilde{\sigma}_3&=& \left( (1+\epsilon)(\frac{M_\pi}{3\pi F_\pi})^2\right)^2 \{ \left[ \frac{40m^2a(s)}{a(s)^2+M_\sigma^2\Gamma_\sigma^2}-1\right]^2 + \frac{(40m^2M_\sigma \Gamma_\sigma)^2} {\left[a(s)^2 + M_\sigma^2\Gamma_\sigma^2 \right]^2} \}
\end{eqnarray}
where $a(s)=M_\sigma^2-s=M_\sigma^2-4M_\pi^2(1+\epsilon)$. This expression consists of two
parts. The first part contains the contributions from the box diagram and from the real part
of the $\sigma$-pole diagram(the expression in the square brackets). The second part
corresponds to the contribution from the imaginary part of the $\sigma$-pole diagram. Both
parts have the common small factor $\delta$
\begin{eqnarray}
\delta&=&\left( \frac{M_\pi}{3 \pi F_\pi}\right)^4
\end{eqnarray}
Now let us compare the behavior of $\tilde{\sigma}_1$ and $\tilde{\sigma}_3$ in vacuum and in
the a dense medium near the two-pion threshold $s=0.1 GeV^2$. In both cases the values of
$\tilde{\sigma}_1$ change very little:
\begin{eqnarray}
\tilde{\sigma}_1 \sim 12-15
\end{eqnarray}
The opposite situation takes place for $\tilde{\sigma}_3$. Indeed, in vacuum we have
$\epsilon=0.28$, $\delta=6.5 \cdot 10^{-4}$ and the main part of $\tilde{\sigma}_3$ is defined
by the contribution connected with the real part of the $\sigma$-pole diagram
\begin{eqnarray}
\left(\frac{40 m^2a}{a^2+M_{\sigma}^2\Gamma_{\sigma}^2}\right)^2 \approx 84
\end{eqnarray}
We can see that the contribution from $\tilde{\sigma}_3$ is very small as compared with the
contribution of the Born term in this case.

A different situation can take place in hot and dense matter. Indeed, at $T=100 MeV$ and
$\mu=290 MeV$ we have $m=138MeV$, $M_\pi=156MeV$, $M_\sigma \approx 317 MeV$, $F_\pi=57MeV$
\cite{ktc}. In this case, the imaginary part of the $\sigma$-pole diagram gives a dominant
contribution to $\tilde{\sigma}_3$. The parameter $\delta$ in this case equals to $7 \cdot
10^{-3}$, and the main contribution from the imaginary part has the form($\epsilon=0.02$)
\begin{eqnarray}
\left( \frac{40 m^2}{M_{\sigma} \Gamma_\sigma} \right)^2 \approx 10^4
\hspace{0.2cm},\hspace{0.5cm} (a \ll M_{\sigma} \Gamma_\sigma)
\end{eqnarray}
As a result, the contribution from the $\sigma$-pole diagram becomes comparable with the
contribution from the Born term.

This effect plays an especially important role for the neutral pion. In vacuum, the cross
section $\pi^0 \pi^0 \rightarrow \gamma \gamma$ is very small. However, in hot and dense
matter the cross sections of $\pi^+ \pi^- \rightarrow \gamma \gamma$ and $\pi^0 \pi^0
\rightarrow \gamma \gamma$ can be comparable.

After qualitative estimations, let us give more exactly the numerical calculation for the
above-considered cases.

The value $s=0.1 GeV^2$ corresponds to the energy of outgoing photons $\omega_\gamma=160 MeV$.
In vacuum ($T=0,\mu=0$), we have the following values for masses $m$, $M_\sigma$ and
parameters $\epsilon$, $\kappa$: $m=280MeV$, $M_\sigma=580MeV$, $\epsilon=0.28$,
$\kappa=0.46$. Using these values, we obtain for the charge pions $\tilde{\sigma}_1 \approx
12$, $\tilde{\sigma}_2 \approx 1.8$, $\tilde{\sigma}_3 \approx 0.1$. For the neutral pion we
have $\tilde{\sigma}_3 \approx 0.04$.

The cross-sections for processes $\sigma_{\pi^+\pi^-\rightarrow \gamma \gamma }$,
$\sigma_{\pi^0\pi^0\rightarrow \gamma \gamma}$ are approximately
\begin{eqnarray}
\sigma_{\pi^+\pi^-\rightarrow \gamma \gamma } \approx 4.8 \mu b,\sigma_{\pi^0\pi^0\rightarrow
\gamma \gamma } \approx 0.015 \mu b
\end{eqnarray}

Now let us consider these processes in hot and dense matter when $T=100MeV$, $\mu=290MeV$.
Here the parameters $\epsilon$, $\kappa$ are equal to $0.02$, $0.14$, respectively. The masses
$m$, $M_\pi$, $M_\sigma$ and value of $F_\pi$ are given above. After substituting the
parameters and masses into (\ref{partsigma}), we obtain
\begin{eqnarray}
\tilde{\sigma}_1 \approx 15.5,\tilde{\sigma}_2 \approx 2.83,\tilde{\sigma}_3 \approx 57.4
\end{eqnarray}
for the charge pions and
\begin{eqnarray}
\tilde{\sigma}_3 \approx 57.5
\end{eqnarray}
for the neutral pions. For the cross-sections $\sigma_{\pi^+\pi^-\rightarrow \gamma \gamma }$,
$\sigma_{\pi^0\pi^0\rightarrow \gamma \gamma}$ we have
\begin{eqnarray}
\sigma_{\pi^+\pi^-\rightarrow \gamma \gamma } \approx 75.6 \mu b,\sigma_{\pi^0\pi^0\rightarrow
\gamma \gamma } \approx 57.5 \mu b
\end{eqnarray}
So, we see that in this domain of $T$ and $\mu$ near the two-pion threshold, $\sigma_3$
increases dramatically. As a result, the charged pions cross-section increases approximately
one and a half order. The neutral pions cross-section increases more than three orders and
becomes comparable with charged pions cross-section.

\begin{figure}
\begin{center}
 \epsfig{figure=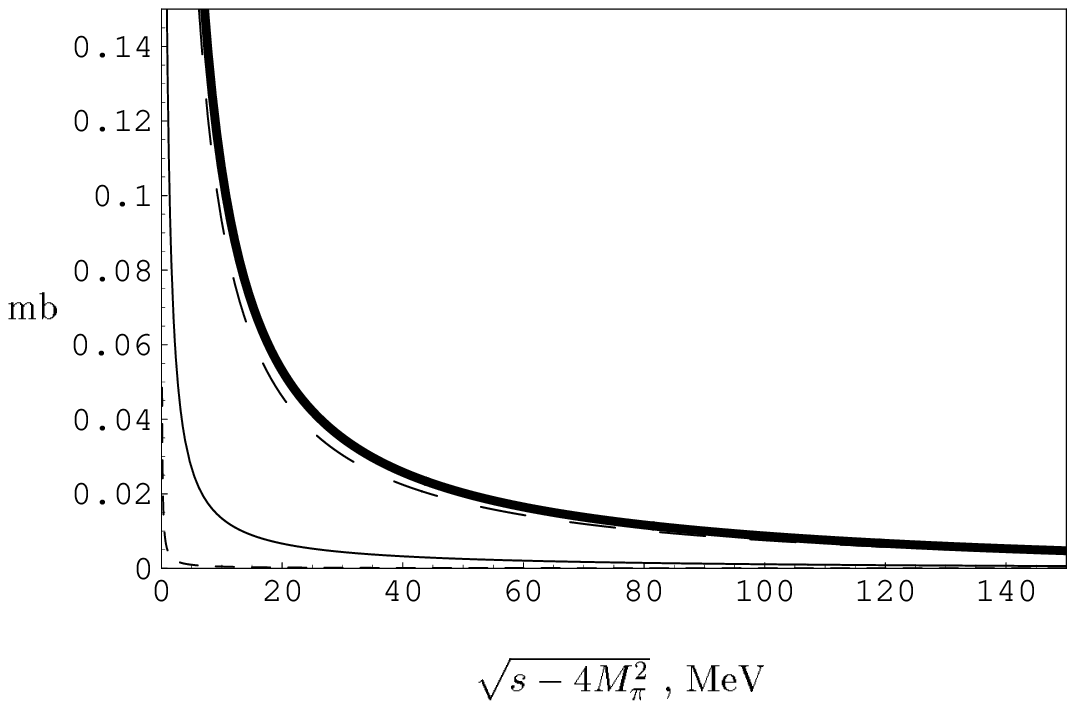,height=8cm,} \caption{}
 \label{crossxcv}
\end{center}
\end{figure}

\begin{figure}
\begin{center}
 \epsfig{figure=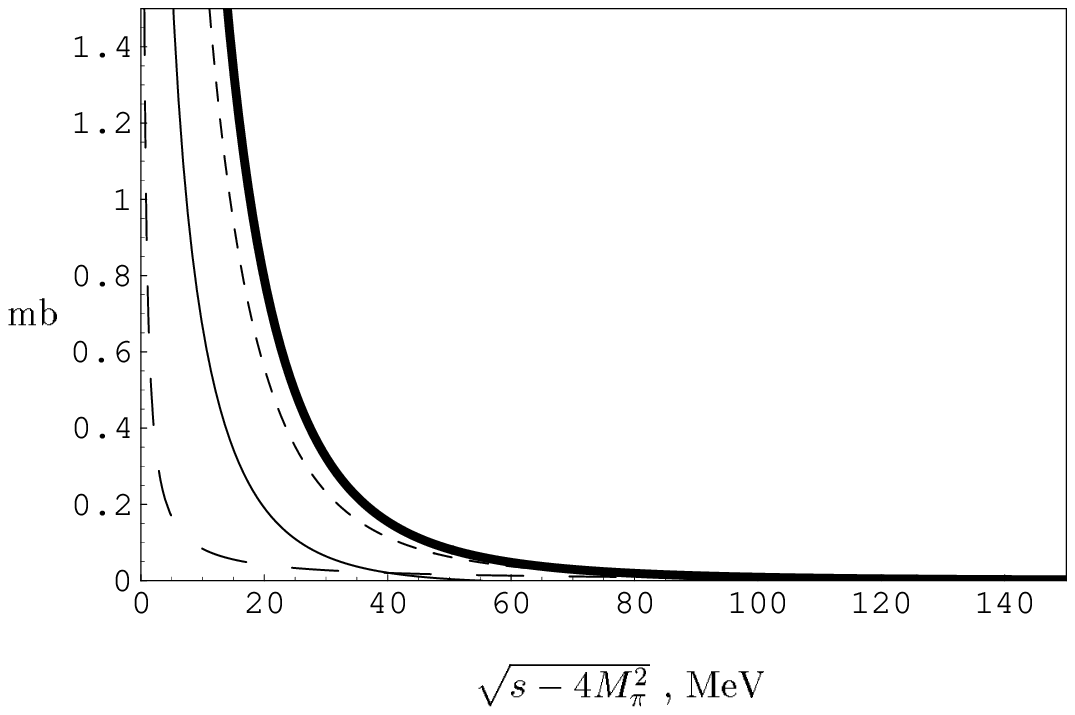,height=8cm} \caption{}
 \label{crossxcm}
\end{center}
\end{figure}

\begin{figure}
\begin{center}
 \epsfig{figure=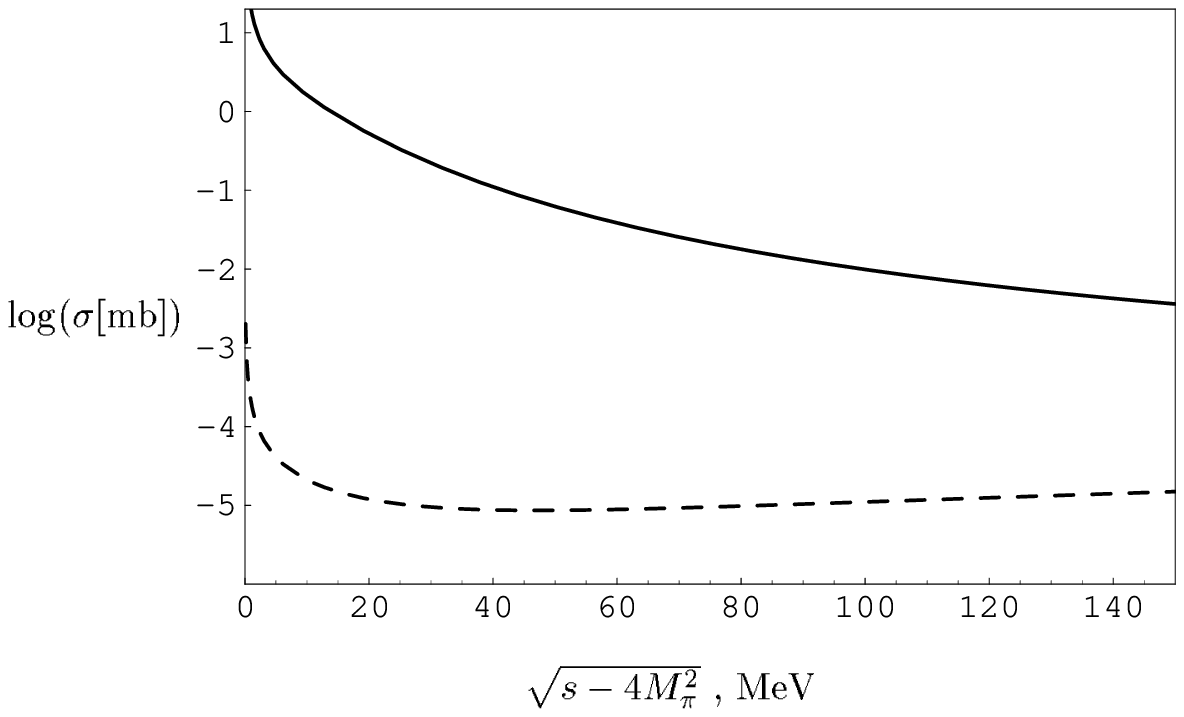,height=7cm} \caption{}
 \label{cross0}
\end{center}
\end{figure}

The cross-sections $\sigma$ and $\sigma_1$, $\sigma_2$, $\sigma_3$ ($\sigma_i=\frac{\pi
\alpha^2}{4s \kappa }\tilde{\sigma}_i$, $i=1,2,3$) for process $\pi^+\pi^-\rightarrow
\gamma\gamma$ are plotted in Fig.\ref{crossxcv} and Fig.\ref{crossxcm} as a function of $s$.
Fig.\ref{crossxcv} shows the numerical results for vacuum where the main contribution comes
from the Born terms. The situation changes at a finite temperature and chemical potential. As
is shown in Fig.\ref{crossxcm}, the $\sigma$-pole diagram gives the dominant contribution and
the cross-section is strongly increased at the threshold. The behavior of the
$\pi^0\pi^0\rightarrow \gamma\gamma$ cross-section is shown in Fig.\ref{cross0} in vacuum and
in hot and dense matter. It is easy to see from Fig.\ref{cross0} that the contribution from
the $\sigma$-pole diagram dramatically increases near the two-pion threshold.

Here we used the approximative expression for the amplitudes where the $T$ and $\mu$
dependence of quark loops was neglected. A more careful calculation was made in
works\cite{VKur,VKur2}.

\section{Discussion and conclusion}

We have shown that the scalar $\sigma$-meson plays an important role in low-energy meson
physics. The $\sigma$-pole diagram gives the main contribution to the $\pi \pi$-scattering
amplitude and ensures its chiral invariance\cite{echaya86,pppp1,pppp2}. The intermediate
$\sigma$-meson gives a dominant contribution to charge pion polarizability in the process
$\gamma\pi\rightarrow\gamma\pi$ \cite{echaya86,pionpolar}. Using the $\sigma$-pole diagram, we
can explain the rule $\Delta I=1/2$ in kaon decays\cite{kspp1,kspp2}. The diagram with the
$\sigma$-meson determines the value of the pion-nucleon
$\Sigma$-term\cite{pns1,actaphys,pns2}. It is not a complete list of significant physical
results where taking the $\sigma$-meson into account allows us to correctly describe hadron
properties in vacuum\footnote{It is worth noticing that to describe the decay $\eta
\rightarrow \pi^0 \gamma \gamma$ in agreement with experiment it is also necessary to take
into account the channel with the intermediate scalar-isovector meson $a_0$(980) }.

However, the behavior of the $\sigma$-meson is especially interesting in a hot and dense
medium. Here the $\sigma$-meson can become a sharp resonance in the vicinity of the critical
values of $T$ and $\mu$. This situation leads to a strong increase of the processes where the
$\sigma$-meson participates as an intermediate particle. In this work, we have demonstrated
this effect on the basis of the process $\pi\pi \rightarrow \gamma\gamma$ (see
\cite{VKur,VKur2}). Similar results were obtained in works \cite{ppgg1,ppgg2}. An analogous
situation can also occur in the $\pi\pi$-scattering process.

The spectral function of a $\pi\pi$-system in the $\sigma$-channel has been studied for finite
densities in \cite{kunih1,kunih2,kunihpp}. Characteristic enhancement of the spectral function
near the two-pion threshold is found. This effect is close to our results obtained for the
process $\pi\pi\rightarrow\gamma\gamma$ (see sec.4).

In work \cite{kunihpp}, T. Kunihiro pointed the first experimental support of his theoretical
results. The CHAOS collaboration\cite{chaos,chaos2} studied the differential cross-sections
$M^A_{\pi\pi}=d \sigma^A/d M_{\pi\pi}$ for the $\pi A \rightarrow \pi^+ \pi^\pm A'$ reaction
on nuclei $A=2,12,40,208$. The observable composite ratio
\begin{equation}
C^A_{\pi \pi}=\frac{ M^A_{\pi\pi} \cdot \sigma^N_{tot}}{M^N_{\pi\pi} \cdot \sigma_{tot}^A}
\label{cr}
\end{equation}
was chosen to disentangle the acceptance issue and because it is slightly dependent on the
reaction mechanism and nuclear distortion (here $\sigma^N_{tot}$ and $ \sigma^A_{tot}$ are the
total cross-sections). It was found out that the $C^A_{\pi^+ \pi^-}$ distribution peak at the
$2m_\pi$ threshold and their yield increase with {\it A}. At the same time, the $C^A_{\pi^+
\pi^+}$ distribution weakly depends on {\it A}. This means that nuclear matter weakly affects
the $(\pi\pi)_{I,J=2,0}$ interactions, whereas the $(\pi\pi)_{I=J=0}$ state forms a strongly
interacting system (the intermediate $\sigma$-meson). The experimental results were compared
with some theoretical models, and the best agreement was found with the models that take into
account the medium modifications of the scalar-isoscalar $\sigma$-meson and the partial
restoration of chiral symmetry in nuclear matter.

The Crystal Ball(CB) collaboration \cite{cb} studied the reaction $\pi^-A \rightarrow
\pi^0\pi^0 A'$ on $H$, $D$, $C$, $Al$, $Cu$. They reported that there was no peak  near the
$2m_\pi$ threshold, observed by CHAOS collaboration, but the increase in strength as a
function of {\it A} was also observed in the $\pi^0 \pi^0$ system. Later, the CB results were
reanalyzed \cite{cbch} in terms of the composite ratio (\ref{cr}) and accounting for different
acceptances of two experiments. It was shown that as far as the $(\pi\pi)_{I=J=0}$ interacting
system is concerned, the results agree with each other very well. This agreement may be
interpreted as an independent confirmation by the CB experiment of the modification of the
$\sigma$-meson properties in nuclear matter, first reported by CHAOS.

The above-described experimental data for $\pi\pi$-scattering allow us to hope that similar
results can be experimentally obtained for the processes $\pi\pi \rightarrow\gamma\gamma$ in
hot and dense matter. It is especially interesting to study $\pi^0\pi^0
\rightarrow\gamma\gamma$, because the cross-section of this process can increase by several
orders of magnitude near the two-pion threshold. Experimental observation of all these effects
will be evidence for approaching the boundary of the domain where the partial restoration of
chiral symmetry and the phase transition of hadron matter into quark-gluon plasma take place.

\begin{center}ACKNOWLEDGMENTS\end{center}

MKV thanks D. Blaschke, E.A. Kuraev, G. R\"opke, S. Schmidt for fruitful discussion and
collaboration. MKV acknowledge support by RFBR(grant no.02-02-16194) and Heisenberg-Landau
program. AER acknowledge support by UNESCO.

\clearpage
\section*{Figure captions}
\begin{enumerate}
\item The quark diagrams describing the $\pi\pi$-scattering. All loops in
Fig.\ref{pipi}-Fig.\ref{2p2g} consist of constituent quarks.
\item The diagram describing the $K_S \rightarrow \pi \pi$ decay with $\Delta I=1/2$. The
dot is a weak vertex.
\item The quark diagrams describing the matrix element $\langle \pi^+(0)\vert {\bar u}u + {\bar d}d\vert \pi^+(0)\rangle$.
The left vertices are scalar quark vertices interacting with a nucleon.
\item The quark diagrams describing the matrix element $\pi \pi \rightarrow \gamma \gamma$.
\item The quark diagrams in the Born approximation describing the matrix element $\pi \pi \rightarrow \gamma \gamma$.
\item The total cross-section $\sigma$(thick solid line) for process $\pi^+\pi^- \rightarrow \gamma\gamma$ and partial cross-sections
 $\sigma_1$(long-dashed line),$\sigma_2$(thin solid line),$\sigma_3$(short-dashed line) in vacuum.
\item The total cross-section $\sigma$(thick solid line) for process $\pi^+\pi^- \rightarrow \gamma\gamma$ and partial cross-sections
 $\sigma_1$(long-dashed line),$\sigma_2$(thin solid line),$\sigma_3$(short-dashed line) in hot and dense matter at $T=100 MeV$, $\mu=290  MeV$.
\item The cross-section for process $\pi^0\pi^0 \rightarrow \gamma\gamma$ in vacuum(dashed
line) and in hot and dense matter $T=100 MeV$, $\mu=290 MeV$(solid line).
\end{enumerate}

\clearpage
\section*{Figures}

\begin{figure}[h]
\setlength{\unitlength}{1mm}
\begin{center}
\begin{picture}(100,40)
\thicklines \put (0,20){\line(1,0){20}} \put (5,23){\bf{$K_S$}} \thinlines \put
(27.5,20){\oval(15,10)} \put (42.5,20){\oval(15,10)} \put (35,20){\circle*{2}} \thicklines
\put (50,20){\line(1,0){20}} \put (60,23){\bf{$\sigma$}} \thinlines \put
(70,20){\line(1,1){10}} \put (70,20){\line(1,-1){10}} \put (80,10){\line(0,1){20}} \thicklines
\put (80,10){\line(1,0){20}} \put (95,13){\bf{$\pi$}} \put (95,33){\bf{$\pi$}} \put
(80,30){\line(1,0){20}}
\end{picture}
\end{center}
\caption{} \label{ksdiag}
\end{figure}

\end{document}